\documentclass[aps,pra,graphicx,9pt,twocolumn,superscriptaddress,showkeys,showpacs]{revtex4-2}
\usepackage[colorlinks,linkcolor=blue,urlcolor=blue,anchorcolor=blue,citecolor=blue]{hyperref}
\usepackage{amsmath}
\usepackage{graphicx}
\usepackage{dcolumn}
\usepackage{mathrsfs}
\usepackage{amssymb}
\usepackage{amsfonts,multirow}
\usepackage{bm,soul,lineno}
\usepackage[final]{changes}

\begin{document}
\title{Experimental observation of spontaneous symmetry breaking in a quantum phase transition}
\author{Wen Ning}
\thanks{These authors equally contributed to the work.}
\author{Ri-Hua Zheng}
\email{ruazheng@gmail.com}
\thanks{These authors equally contributed to the work.}
\author{Jia-Hao L\"{u}}
\author{Fan Wu}
\affiliation{Fujian Key Laboratory of Quantum Information and Quantum
Optics, College of Physics and Information Engineering, Fuzhou University,
Fuzhou, Fujian, 350108, China}
\author{Zhen-Biao Yang}\email{zbyang@fzu.edu.cn}
\author{Shi-Biao Zheng}\email{fjqiqo@fzu.edu.cn}
\affiliation{Fujian Key Laboratory of Quantum Information and Quantum
Optics, College of Physics and Information Engineering, Fuzhou University,
Fuzhou, Fujian, 350108, China}
\affiliation{Hefei National Laboratory, Hefei 230088, China}
\begin{abstract}

Spontaneous symmetry breaking (SSB) plays a central role in understanding a large variety of phenomena associated with phase transitions, such as superfluid and superconductivity. So far, the transition from a symmetric vacuum to a macroscopically ordered phase has been substantially explored. The process bridging these two distinct phases is critical to understanding how a classical world emerges from a quantum phase transition, but so far remains unexplored in experiment. We here report an experimental demonstration of such a process with a quantum Rabi model engineered with a superconducting circuit. We move the system from the normal phase to the superradiant phase featuring two symmetry-breaking field components, one of which is observed to emerge as the classical reality. The results demonstrate that the environment-induced decoherence plays a critical role in the SSB.
\end{abstract}

\maketitle

\section{Introduction}
The concept of spontaneous symmetry breaking (SSB) has infiltrated almost
all research branches of modern physics, ranging from condensed physics \cite{sachdev2011}
to quantum field theory \cite{Nambu2009,Weinberg2004}. SSB appears when the observed systems lack
the symmetry of the governing physical laws. It underlies phase transitions
described by the Ginzburg-Landau-Wilson paradigm, e.g.,
Bardeen-Cooper-Schrieffer superconducting phenomena \cite{Bardeen1957}, and is critical to the Higgs mechanism \cite{Bernstein1974}, where the mass is generated by breaking the invariance associated with the Lagrangian.
Besides these, SSB plays an important role
in the evolution of biology, e.g., the visual cortex processes image by
spontaneously breaking both the translation and rotation symmetries, which
was recently interpreted in terms of zero-temperature phase transitions \cite{Fumarola2022}.
In a more general sense, SSB is responsible for the formation of the
universe \cite{kibble1980}, and thus gives arise to the infinite diversity of nature.

In quantum field theory, SSB occurs in an
infinitely extended system that possesses degenerate vacua. In this case, an
infinitesimal perturbation is sufficient to single out one of these vacua as
the real ground state. For a quantum system with a definite size, the
symmetry is generally broken by the decoherence process \cite{van2005,dziarmaga2012,Pintos2019}, where the
environment or a monitoring meter continuously gathers the information about
the system by entangling its state with the system. This process progressively destroys the quantum coherence between the symmetrically superposed quasiclassical
components with broken symmetry, transforming the quantum superposition
into a classical mixture \cite{zurek1991}. The decoherence rate linearly scales with the
size of the system, which prohibits observation of quantum superpositions in
a macroscopic object, such as the famous Schr\"{o}dinger cat \cite{schrdinger1935}. In this
case, the environment or meter plays a central role in the symmetry breaking of the observed system.

Despite impressive experimental demonstrations in classical systems \cite{liu_spontaneous_2014, hamel_spontaneous_2015, PhysRevLett.118.033901, xu_spontaneous_2021,PhysRevLett.128.053901}, observations of the SSB associated with quantum phase transitions have remained an outstanding task.
As the energy gap vanishes at the critical point, an adiabatic
evolution across this point would require an infinitely long time. For this
reason, experimental demonstrations of quantum phase transitions have been focused
on out-of-equilibrium systems  \cite{Cai2021,Safavi2018,feng2015,Fink2017,zheng2022,zhang2017,Jurcevic2017,xu2020,klinder2015,zhiqiang2017,baumann2010,brennecke2013,Baumann2011,leonard2017nature,leonard2017science,Ferri2021,zhang2021}, where transitions between distinct
phases were realized by a quenching process or in a dynamical manner. In
thus-realized superradiant phase transitions (SPTs) with an ultracold atomic gas
trapped in an optical cavity, the breaking of both discrete \cite{Baumann2011} and continuous
translation symmetries \cite{leonard2017nature,leonard2017science} have been observed. However, in these
experiments, the symmetries were not broken spontaneously, but explicitly
with a small symmetry-breaking field or by a detuning that broke the
Hamiltonian symmetry. In two very recent experiments, continuous symmetry-breaking phases were prepared in a one-dimensional ionic chain \cite{feng2022} and a
two-dimensional Rydberg array \cite{chen2023}, respectively.
Despite these impressive progresses, it remains an experimental challenge to directly observe how the SSB occurs during a quantum phase transition, which is responsible for the emergence of a classical, symmetry-breaking state under a symmetric Hamiltonian.
\section{Physical Model and Theoretical Predictions}
We here report an experimental observation of such a process in an effective
Rabi model, engineered in a circuit quantum electrodynamics (QED)
architecture. With the high level of control over the effective coupling
between the qubit and the photonic field involved in the Rabi Hamiltonian,
we realize a second-order SPT, where the system is moved from a normal phase (NP) to a superradiant phase
(SP) formed by two coherent field states correlated to different qubit
states. We observe the progressive evolution of the field from the initial
symmetric vacuum to two superposed components, each of which has a broken
symmetry in phase space, and the transition of their symmetric superposition
to a classical mixture during their growth. Our results reveal that SSB is
closely related to a quantum natural selection made by the environment \cite{lloyd2009},
which continuously performs measurements on the system but with
inaccessible records, washing the quantum coherence between the two
quasi-classical components emerging from a quantum phase transition.

The system under investigation is comprised of a qubit and a photonic field,
coupled by the Rabi Hamiltonian \cite{Ashhab2013,Hwang2015}%
\begin{equation}
H_{RM}=\delta a^{\dagger }a+\frac{1}{2}\Omega \sigma _{z}+\frac{1}{2}\lambda
\sigma _{x}(a^{\dagger }+a).
\end{equation}
Here $a^{\dagger }$ and $a$ denote the creation and annihilation operators
for the photonic field with a frequency $\delta $, $\sigma _{x}=\left\vert
e\right\rangle \left\langle g\right\vert +\left\vert g\right\rangle
\left\langle e\right\vert $ and $\sigma _{z}=\left\vert e\right\rangle
\left\langle e\right\vert -\left\vert g\right\rangle \left\langle
g\right\vert $ are pseudo-spin operators for the qubit, which has a
frequency $\Omega $, and whose upper and lower levels are denoted as $%
\left\vert e\right\rangle $ and $\left\vert g\right\rangle $. The qubit-resonator coupling strength is $\lambda $. The Rabi Hamiltonian is invariant
under the parity transformation $(-1)^{a^{\dagger }a+|e\rangle \langle e|}$, featuring a Z$%
_{2}$ symmetry. In the limit $\Omega /\delta \rightarrow \infty $, the
system has a unique ground state below the critical point $\xi =\lambda /%
\sqrt{\Omega \delta }=1$, given by $\left\vert \psi _{np}\right\rangle
=S(r_{np})\left\vert 0\right\rangle \left\vert g\right\rangle $, where $%
S(r_{np})$ is a squeezing operator with $r_{np}=-\frac{1}{4}\ln (1-\xi ^{2})$%
. Due to the limited frequency ratio $\Omega /\delta $ in a real experiment,
when $\xi <1$, the field in the experimentally realized Rabi model
almost remains in the vacuum state, which is
significantly distinct from the ideal ground state for $\Omega /\delta
\rightarrow \infty $. Above the critical point, the system enters the SP,
characterized by two degenerate ground states $\left\vert \psi _{sp}^{\pm
}\right\rangle =D(\pm \alpha )S(r_{sp})\left\vert 0\right\rangle \left\vert
\pm \right\rangle $, where $r_{sp}=-\frac{1}{4}\ln (1-\xi ^{-4})$, $D(\alpha
)=\exp [\alpha (a^{\dagger }-a)]$ with $\alpha =\sqrt{\frac{\Omega }{4\xi
^{2}\delta}(\xi ^{4}-1)}$, and $\left\vert \pm \right\rangle $ are two
qubit basis states. \added{Compared with the case in the Dicke model, the SPT in the Rabi mode is realized by replacing thermodynamical limit with the scaling limit, where the frequency ratio between qubit and the field tends to infinity \cite{Hwang2015}.} Each of these ground states is symmetry-broken,
signified by the nonvanishing coherence $\left\langle \psi _{sp}^{\pm
}\right\vert a\left\vert \psi _{sp}^{\pm }\right\rangle =\pm \alpha $ \cite{Hwang2015}. Under
unitary Hamiltonian dynamics, adiabatically increasing the parameter $\xi $
across the critical point would evolve the system to the equal superposition
of the two degenerate ground states%
\begin{equation}
\left\vert \psi _{e}\right\rangle =N_{e}(\left\vert \psi
_{sp}^{+}\right\rangle +\left\vert \psi _{sp}^{-}\right\rangle ),
\end{equation}
which preserves the parity of the ground state below the critical point.
However, during the evolution, the system is inevitably coupled to the
environment, which continuously monitors the system in the preferred basis \cite{zurek1991}, forcing the system to choose between these two conflicting classical states.

\section{Device and experimental scheme}
The experiment is performed
with a circuit QED device, which involves a resonator of a fixed frequency $\omega _{r}/2\pi =5.581$ GHz controllably coupled to 5 frequency-tunable Xmon qubits, with an on-resonance
coupling strength of $g/2\pi \simeq 20$ MHz. One of these qubits is used as the test qubit
for realizing the Rabi model, while the other qubits are effectively decoupled from the resonator
during the quenching process due to the large detunings. The experiment
starts by tuning the test qubit to the frequency $\omega _{q}/2\pi= 5.21$ GHz, where it
is subjected to a transverse microwave driving of an amplitude $A=2\pi\times 15$ MHz and two
longitudinal parametric modulations, one having a fixed frequency $\nu _{1}/2\pi= 185$ MHz
and amplitude $\varepsilon _{1}= 2\pi\times 146$ MHz, while the other bearing a fixed frequency
$\nu _{2}/2\pi= 25.97$ MHz but a tunable amplitude $\varepsilon _{2}$. With this setting,
the qubit is coupled to the transverse field at the carrier and to the
resonator at the second upper sideband with respect to the first modulation
\cite{zheng2022}. In the framework coinciding with the Rabi precession associated with
the transverse drive and discarding the fast rotating terms, the system
dynamics can be described by the effective Rabi Hamiltonian with an
effective coupling strength $\lambda =g J_{2}(\mu )/2=2\pi\times0.735$ MHz, where $%
J_{n}(\mu )$ is the $n$th Bessel function of the first
kind with $\mu =\varepsilon _{1}/\nu _{1}$. The effective qubit frequency
and resonator frequency are $\Omega =\varepsilon _{2}/2$ and $%
\delta =\omega _{r}-\omega _{q}-2\nu _{1}$, respectively. The quench
parameter $\xi $ is controllable by adjusting the amplitude $\varepsilon
_{2} $ of the second parametric modulation and by tuning the Stark shift of
the resonator produced by an ancilla qubit dispersively coupled to the
resonator.

The quench parameter $\xi $ is modeled as a linear function of time, $\xi =0.5+2t/(3 \ \mu s)$.
We first investigate the evolution of the resonator irrespective of the
qubit state. To monitor the real-time evolution of the resonator, after a
preset quench time $t$ we switch off both the transverse and longitudinal
drives, decoupling the test qubit from the resonator. Then the Wigner tomography of the
resonator's state is performed by applying a microwave pulse to the resonator
to produce a displacement operation $D(-\beta )$, tuning an ancilla qubit
on resonance with the resonator for a given time, and then biasing it back to
its idle frequency for state readout. The Wigner function and the reduced
density operator $\rho _{r}$ for the resonator can be extracted from the
recorded Rabi oscillations signals of the ancilla qubit.
\section{loss-induced SSB}
The classical characteristic of the field can be well characterized by the
Q-function, defined as, $Q(\gamma )=\frac{1}{\pi }\left\langle \gamma
\right\vert \rho _{r}\left\vert \gamma \right\rangle $, where $\left\vert
\gamma \right\rangle $ is the coherent state with a complex amplitude $%
\gamma $. The resonator's Q-functions, measured for quench times $t=0$, $%
2.0$, $2.5$, and $2.8$ $\mu $s, are displaced in Fig. 1a. As expected, the Q
function gradually evolves from the single- to double-peak structure during
the quenching process. Each of these two peaks corresponds to a quasi-classical
state with a broken symmetry, manifested by the nonzero phase-space
coordinates of the local maxima, $\alpha _{+}$ and $\alpha _{-}$,
which can serve as the order parameter \cite{Hwang2016}. The two peaks correspond to quasi-classical coherent states $|\pm \alpha\rangle$
with complex amplitudes $\pm \alpha$.
The two-peak pattern represents a signature of the second-order phase transition without coexistence of the normal and superradiant phases. This phase transition is in distinct contrast with the first-order SPT that features emergence of three peaks \cite{Zhu2022}, one of which corresponds to the NP that coexists with the two SP components, as demonstrated in the previous experiment \cite{zheng2022}. This significant distinction is
mainly due to the fact that the control parameter $\xi $ is linearly
increased in the present experiment, but scaled in a highly nonlinear manner
in the previous experiment. \added{Due to the limitation of the experimental frequency ratio, $\Omega / \delta = 10$, the photon number does not show a divergent behavior at the critical point $\xi=1$, corresponding to $t=0.75\ \mu$s, as shown in Fig. S3. This is due to the fact that the strongly squeezed resonator state, predicted to emerge near the critical point for the infinite frequency ratio \cite{Hwang2015}, does not appear in the experiment.}

\begin{figure}
\includegraphics[width=8.4cm]{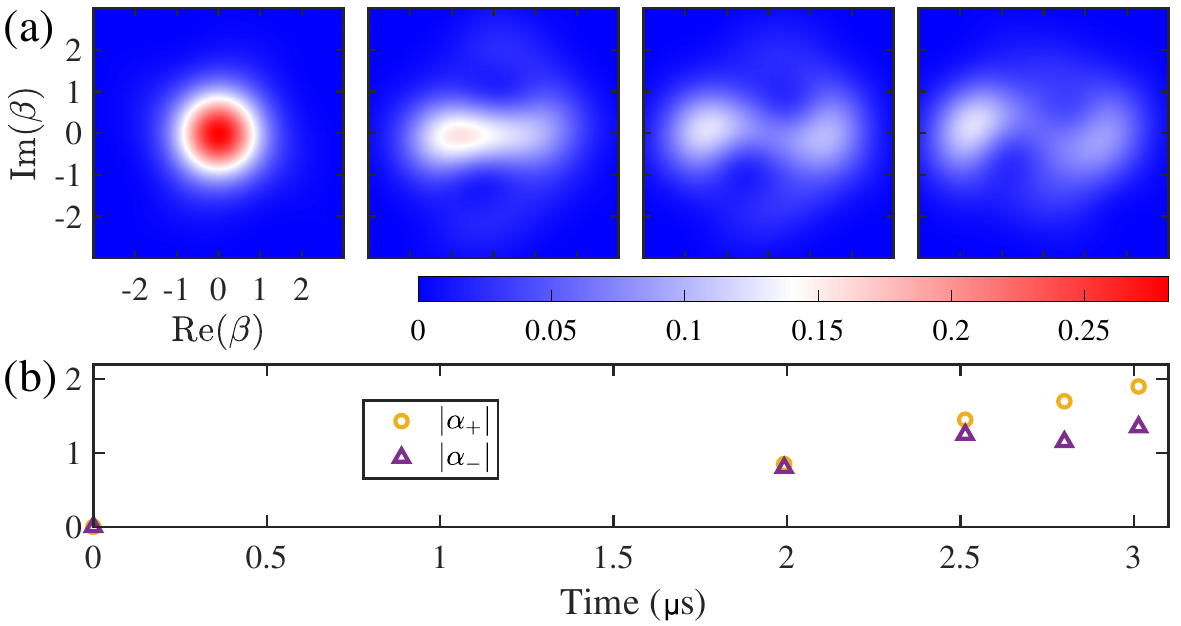}
\caption{\textbf{Observation of phase-space evolution during the SPT}.
(a) Q-functions, measured for quench times $t=0$, $2.0$, $2.5$, and $2.8$ $\mu$s. During the
process, the quench parameter as a function of time is modeled as $\xi =0.5+2t/(3 \ \mu s)$.
After the interaction between the test qubit and the resonator is
effectively switched off, the resonator state is read out by subsequently
performing a phase-space translation to the resonator, resonantly coupling
it to an ancilla qubit, and recording the Rabi signals of the ancilla. (b)
Phase-space coordinates of the local maxima, $\alpha _{+}$ and $%
\alpha _{-}$, versus quench time.}
\label{fig1}
\end{figure}

Figure. 1b displays the measured $\left\vert
\alpha _{+}\right\vert $ (dots) and $\left\vert \alpha _{-}\right\vert $
(triangles) versus time. \replaced{Due to the experimental imperfections, there is a certain degree of unbalance between the two degenerate ground states, which is continually enlarged when the control parameter $\xi$ is increased.}{Due to the experimental imperfections, after the
time $t \simeq2.0$ $\mu$s $\left\vert \alpha _{+}\right\vert $ and $\left\vert \alpha
_{-}\right\vert $ become inconsistent with each other and the discrepancy is
continually enlarged.} \replaced{This discrepancy manifests the explicit symmetry breaking owing to the deviation of the realized Hamiltonian dynamics from the ideal symmetric case, which, however, does not destroy the quantum coherence between the two superimposing ground states.}{This discrepancy, as well as the unbalanced
populations of these two components, manifests the explicit symmetry
breaking owing to the deviation of the realized Hamiltonian dynamics from
the ideal symmetric case, which is beyond the scope of the present work.} We
here focus on the SSB process where one of the two growing coherent states
is singled out as the classical reality by the environment, which
continuously gathers the information about which classical state the system
is in, thereby deteriorating the quantum coherence that makes the two
classical states suspending.
The Q-functions of the resonator do not manifest the quantum coherence
between the classical components. To obtain full information about this
nonclassical behavior, it is necessary to perform a joint qubit-resonator state tomography.
The information of the field state associated with the qubit state $\vert j\rangle~(j=g,e)$ is contained in the corresponding Wigner function, defined as
\begin{equation}
	{\cal W}_{j}(\beta)=\frac{2}{\pi P_j}\sum(-1)^n P_{j,n}(\beta),
\end{equation}
where  $P_{j,n}(\beta)=\langle j,n\vert D(-\beta)\rho D(\beta)\vert j,n\rangle $, and $P_{j}$ is the probability of detecting the qubit in $\vert j\rangle $-state. Here the symbol ``$n$" denotes the photon number in the resonator. These matrix elements are extracted by detecting the test qubit, and correlating the measurement outcome to the Rabi oscillation signals of the ancilla qubit resonantly coupled to the resonator following a displacement operation $D(-\beta)$ \cite{zheng2022,hofheinz2009,Zheng2015}.
Figure. 2a shows the conditional Wigner function, ${\cal W}_{e}(\beta )$, measured after
quench times $t=2.5$, $2.8$, and $3.0$ $\mu $s. The results clearly reveal
a progressive loss of the quantum coherence between the two classical
components, manifested by the gradual washing-out of the interference
fringes between the two Gaussian-like peaks.
Such a process can be understood in terms of the competition between the symmetric Hamiltonian dynamics and naturally-occurring dissipation, which continuously leaks the phase information about the growing field to the environment \cite{Pintos2019,zurek1991}. This natural information acquisition accounts for coherence loss between the two suspending classical states, which leads to the SSB during the quantum phase transition. Due to the imperfect Hamiltonian control, the two Gaussian peaks are gradually diffused during the quench, which limits the size of the accessible output coherent field. The observation of the SSB can be pushed to a larger size with a more deliberately designed and fabricated device.
The other conditional Wigner function ${\cal W}_{g}(\beta)$ measured exhibits a similar evolution behavior, as shown in Fig.
2b. \added{In all the displayed Wigner functions, the main interference fringes appear in a region around the origin, which agrees with the numerical simulation presented in Fig. S6.} It should be noted that the measured Q functions and Wigner functions
present symmetric structures even for a classical mixture, which can be
interpreted as follows. For each experimental implementation, one of the two
symmetry-breaking classical states, but not both, is singled out as the
reality. The symmetry of the phase-space distribution associated with such a
classical mixture reflects that the two conflicting classical realities
appear with the same probability.

\begin{figure}
\includegraphics[width=8.4cm]{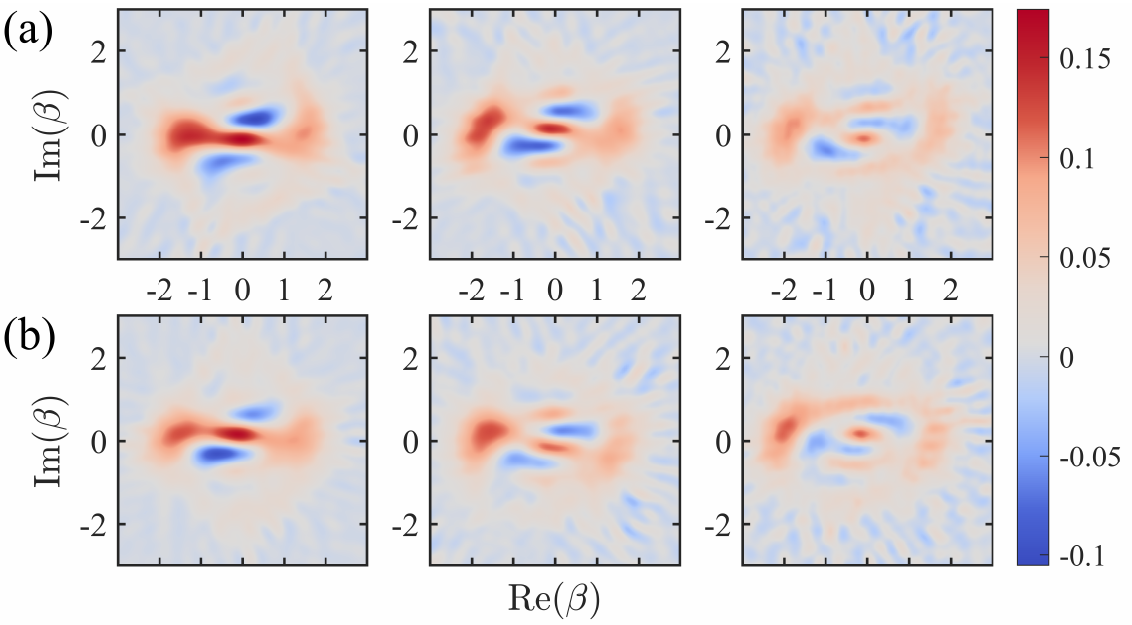}
\caption{\textbf{Observation of the loss of the quantum coherence.} (a) Wigner functions associated with the qubit state $\vert e\rangle $, measured after quench times $t=2.5$, $2.8$, and $3.0$ $\mu$s. The results are obtained by correlating the Wigner function of the resonator to the measurement outcome $\vert e\rangle $ of the test qubit. (b) Measured Wigner functions associated with the qubit state $\vert g\rangle $.}
\label{fig2}
\end{figure}

Previously, the decoherence processes of Schr\"{o}dinger-cat-like quantum superpositions have been
demonstrated in cavity QED \cite{Brune1996,deleglise2008}, ion-trap \cite{myatt2000} and mechanical
resonators \cite{science2023}, where the
cat states were pre-prepared, following which the decoherence effects were
observed. However, in reality, the environment is always playing its role,
instead of waiting to set in until the cat state has grown up. Our
experiment illustrates such a process, where the parity symmetry is broken
as a consequence of the competition between the unitary breeding of the cat
state and the environment-induced dissipation. Such a competition is
critical to the emergence of the classical reality from a quantum world, but
has not been experimentally reported so far. Our results reveal that
decoherence is responsible for breaking the symmetry inherent in the
Hamiltonian dynamics.

The symmetry-breaking process can be further characterized by the evolution of
the nonclassical volume \cite{Anatole2004}, defined as the phase-space integral of the
quasi-probability distribution over the regions with negative values. The nonclassical volume of the field state associated with the qubit state $\vert j \rangle$ is calculated as ${\cal V}_{k}=\left( 1-\int \mathrm{d}^{2}~\beta \left\vert {\cal W}_{k}(\beta )\right\vert\right) /2$
 which quantifies the quantum coherence between the two classical components. Figure. 3 displays ${\cal V}_{e}$ and ${\cal V}_{g}$,
inferred from the Wigner matrices reconstructed for different quench times.
When the phase-space separation between $\left\vert \alpha _{\pm
}\right\rangle $ is small, the unitary dynamics dominate over the
environment-induced decoherence, so that the quantum coherence is improved
with the quench time. With the growth of the photon number, the cat state
becomes increasingly vulnerable to decoherence. Finally, decoherence
plays a dominant role, destroying the quantum coherence between the two
classical components, and manifesting SSB.

\begin{figure}
\includegraphics[width=8cm]{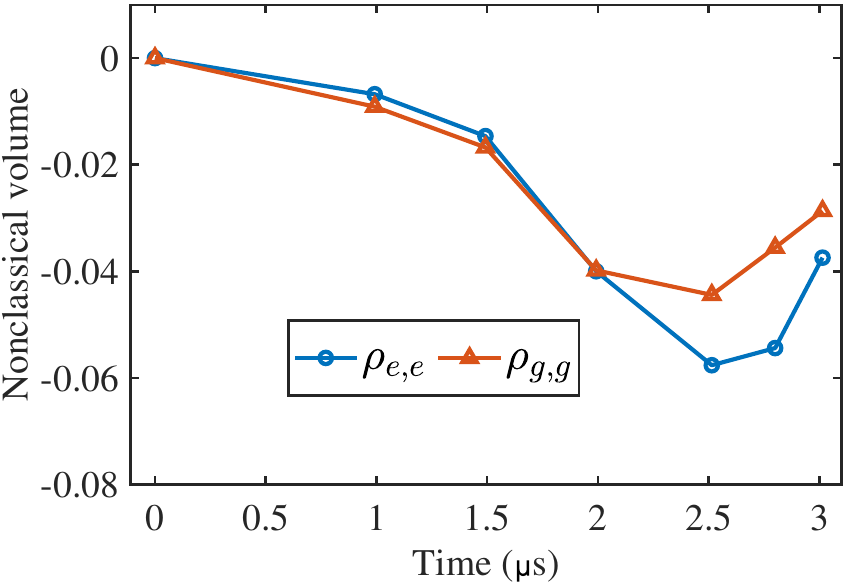}
\caption{\textbf{Measured nonclassical volumes (${\cal V}_{k}$) for different the quench time.} The nonclassical volume of the resonator's output state associated with the qubit state $\left\vert k\right\rangle $ ($k=e,g$) is defined as the integral of the corresponding Wigner function ${\cal W}_{k}(\beta )$ over the regions in phase space where ${\cal W}_{k}(\beta )$ is negative. As the negativity is a consequence of the phase-space quantum interference between the two classical components, the evolution of ${\cal V}_{k}$ manifests the process during which how the quantum coherence grows and then drops under the competition between the Hamiltonian dynamics and decoherence.}
\label{fig3}
\end{figure}
\section{Conclusion}
In conclusion, we have observed the spontaneous symmetry-breaking process in the
second-order phase transition, realized with an effective Rabi model
synthesized in a superconducting circuit. With a high level of control over the
system parameters, we move the system from the NP to a
SP with two symmetry-breaking classical components, each
corresponding to a photonic coherent state with a well-defined amplitude and
phase. We monitor the SSB process, during which one of these two components
is singled out as the classical reality. The results build a close relationship
between SSB and environment-induced decoherence. With the improvement of the coherence times of the system,
a SSB process can be observed with a larger cat size, which would shed new light on the quantum-to-classical transition.
\section{Acknowledgements}
\textbf{Funding:} This work was supported by the National Natural Science Foundation of China
under (Grant No. 11874114, Grants No. 12274080,
and No. 11875108) and Innovation Program for Quantum Science and
Technology under Grant No. 2021ZD0300200.
\textbf{Competing interests:} The authors declare that they have no competing interests.
\textbf{Data and materials availability:} All data needed to evaluate the conclusions
in the paper are present in the paper and/or the Supplementary Materials. Additional data related
to this paper may be requested from the authors.
\bibliography{ref}

\end{document}



\title{Supplementary Material for\\ ``Experimental observation of spontaneous symmetry breaking in a quantum phase transition''}

\author{Wen Ning}
\thanks{These authors equally contributed to the work.}
\author{Ri-Hua Zheng}
\email{ruazheng@gmail.com}
\thanks{These authors equally contributed to the work.}
\author{Jia-Hao L\"{u}}
\author{Fan Wu}
\affiliation{Fujian Key Laboratory of Quantum Information and Quantum
Optics, College of Physics and Information Engineering, Fuzhou University,
Fuzhou, Fujian, 350108, China}
\author{Zhen-Biao Yang}\email{zbyang@fzu.edu.cn}
\author{Shi-Biao Zheng}\email{fjqiqo@fzu.edu.cn}
\affiliation{Fujian Key Laboratory of Quantum Information and Quantum Optics, College of Physics and Information Engineering, Fuzhou University, Fuzhou, Fujian, 350108, China}
\affiliation{Hefei National Laboratory, Hefei 230088, China}
\date{\today}
\maketitle
\tableofcontents

\section{Derivation of the effective Hamiltonian}
The physical model considered is composed of a photonic field coupling to a qubit, whose frequency is modulated according to $\omega_{q}(t) = \omega_{0} + \varepsilon_{1}\cos(\nu_{1}t)+\varepsilon_{2}\cos(\nu_{2}t)$. Besides, the qubit is transversely driven by an external field at the frequency $\omega_0$ with an amplitude $A$, the Hamiltonian in the rotating frame with respective to $H_0  = (\omega_0 + 2\nu_1)a^\dagger a + [\omega_0 + \epsilon_1 \cos(\nu_1 t)] \sigma_z/2$ is ($\hbar=1$ hereafter)
\begin{eqnarray}\label{eqS1}
    H_{I}^{\prime }=\delta a^{\dagger }a+\frac{1}{2}\varepsilon _{2}\cos (\nu
    _{2}t)\sigma_{z}+\left\{\exp \left[ -i\mu \sin (\nu _{1}t)\right] \left[g \exp
    \left( 2i\nu _{1}t\right) a^{\dagger }+A \right]\sigma_{-} +{\rm{H.c.}}\right\}.
\end{eqnarray}%
Here, $\mu =\varepsilon _{1}/\nu _{1}$, $\delta$ is the effective resonator frequency, $g$ is the qubit-resonator coupling strength measured on their resonance, $\omega_0$ corresponds to the mean transition frequency of the qubit, and $\varepsilon_j$ ($\nu_j$) is the $j$th modulation amplitude (frequency). The above Hamiltonian can be expanded by the Jacobi-Anger expansion as

\begin{eqnarray}\label{eqS6}
    H_{I}^{\prime } &=&\delta a^{\dagger }a+\frac{1}{2}\varepsilon _{2}\cos (\nu
    _{2}t)\sigma_{z} + \sum\limits_{m=-\infty }^{\infty}H_m,\\ \nonumber
    H_m &=& J_{m}(\mu )\left\{g \exp [-i(m-2)\nu _{1}t]a^{\dagger }+A \exp \left(
    -im\nu _{1}t\right) \right\}\sigma_{-}
    +{\rm{H.c.}}
\end{eqnarray}
Under the condition $\left\{\left\vert g J_{2}(\mu )\right\vert,~\left\vert A J_{0}(\mu )\right\vert,~\delta,~\varepsilon_2\right\} \ll \nu_{1}$, the fast-oscillating terms can be discarded. In such a case, $H_{I}^{\prime }$ reduces to
\begin{eqnarray}
    H_{I}^{\prime } &=& \delta a^{\dagger }a+\frac{1}{2}\varepsilon _{2}\cos (\nu_{2}t)\sigma_{z} + [gJ_{2}(\mu) a^{\dagger } +AJ_{0}(\mu)]\sigma_{-} +{\rm{H.c.}}  \\ \nonumber
    &=&\delta a^{\dagger }a+\frac{1}{2}\varepsilon _{2}\cos (\nu_{2}t)\sigma_{z} + \frac{1}{2}B_{0}\sigma _{x}+\eta \left(X\sigma _{x}-Y\sigma _{y}\right),
\end{eqnarray}
where $B_{0}=2AJ_{0}(\mu )$, $\eta =g J_{2}(\mu )/2$, $X=a^{\dagger
}+a $, $Y=i(a^{\dagger }-a)$, $\sigma_{x}=\sigma_{-}+\sigma_{-}^{\dag}$, and $\sigma_{y}=i\sigma_{-}-i\sigma_{-}^{\dag}$.
Under the unitary transformation $\exp (iB_0\sigma _{x}t/2)$ and in the limit of $B_0\gg \eta ,\varepsilon _{2}/2$ and $\nu _{2}=B_{0}$, we obtain the effective Rabi Hamiltonian
\begin{eqnarray}\label{eqS9}
    H_{R}=\frac{1}{2}\Omega \sigma _{z}+\delta a^{\dagger }a+\eta \sigma_{x}(a^{\dagger }+a),
\end{eqnarray}%
where $\Omega=\varepsilon_2/2$ is the effective qubit frequency.

\section{Control of the quenching dynamics}
\subsection{Time-dependence of the normalized coupling parameter $\xi$}
\begin{figure}[htb]
    \centering
    \includegraphics[width=14cm]{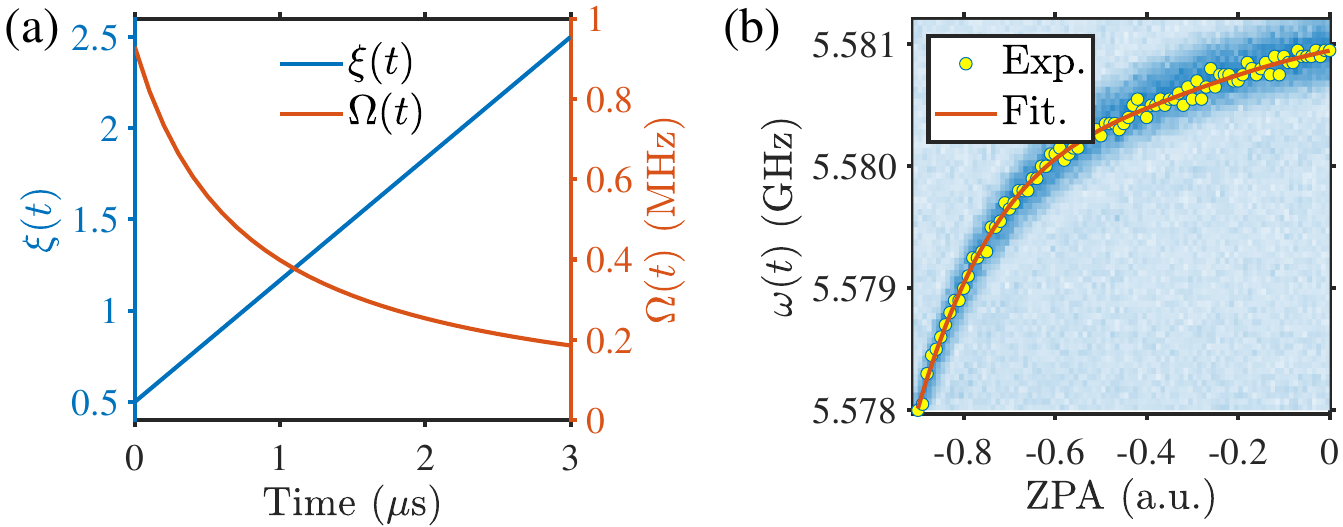}
    \caption{(a) The normalized coupling parameter $\xi$ and the effective qubit frequency $\Omega(t)$ versus time, exhibiting through the left and right y-axes, respectively. The effective resonator frequency $\delta(t)$ is set as $\Omega/10$. (b) The actual resonator frequency $\omega_r(t)$ versus ZPA. The yellow dots represent the experimental data, with the red solid as a fitting.}
    \label{xidwc}
\end{figure}

During the quenching process, we adjust the effective qubit frequency $\Omega$ and the effective resonator frequency $\delta$, so that the normalized coupling parameter $\xi$ slowly passes through the quantum transformation critical point:
\begin{eqnarray}
    \xi&=&\lambda /\sqrt{\Omega \delta} =  \xi_0 + (\xi_{\max}-\xi_0)t/t_f,
\end{eqnarray}
where $\lambda$ is the qubit-resonator coupling strength, $\xi_{\rm max}$ ($\xi_{0}$) is the maximum (initial) value of $\xi(t)$, and $t_{f}$ is the total evolution time. For Fig. 1 and Fig. 2 in the main text, we set $\xi_{0}=0.5$, $\xi_{\rm max}=2.5$ and $t_f = 3\ \mu$s. Note that, due to the fact practically it's hard to reach the limit of $\Omega/\delta \rightarrow \infty$, we choose and keep $\Omega/\delta = 10$ during the whole quenching process to approximate the preset limitation $\Omega \gg \delta $ of the SPTs. The variances of $\xi$ and $\Omega$ over time are plotted and shown in Fig.~\ref{xidwc}a.

\subsection{Control of the effective qubit frequency $\Omega$ and the effective resonator frequency $\delta$}
According to Eq.~(\ref{eqS9}), the effective qubit frequency $\Omega$ can be constructed by adjusting amplitude $\varepsilon_{2}$ of the second parametric modulation, with $\Omega = \varepsilon_{2}/2$. While the effective resonator frequency $\delta$ is controllable by tuning the Stark shifts induced by the dispersive coupling of the resonator to an ancilla qubit, whose frequency is flexibly modulable.

When the detuning $\Delta\omega$ between the ancilla qubit and the resonator is varied from $\Delta\omega_1$ to $\Delta\omega_2$, and remains much larger than the coupling strength $g^{\prime}$ between them, the resulting resonator frequency shift becomes
\begin{eqnarray}\label{r_shift}
    \delta\omega = g^{\prime}(\frac{1}{\Delta\omega_1} - \frac{1}{\Delta\omega_2}).
\end{eqnarray}
In another form, Eq.~(\ref{r_shift}) can be written as
\begin{eqnarray}\label{rr_shift}
    \delta(t)-\delta(0)=\frac{g^{\prime 2}}{f(0)-\omega_r}-\frac{g^{\prime 2}}{f(t)-\omega_r},
\end{eqnarray}
where $g^{\prime}/2\pi=20.91$ MHz, and $\omega_r$ is the fixed resonator frequency measured with all other qubits staying at their ground states and tuned to their idle frequencies.

For simplicity, to accurately control $\delta(t)$, we directly measure the relationship between the actual resonator frequency $\omega_r(t)$ and the z-line pulse amplitude (ZPA) applied in the ancilla qubit, instead of calculating it from Eq.~(\ref{rr_shift}).
We apply a square pulse of a certain amplitude to the Z-line of the ancilla qubit to bias it to the frequency $f(t)$. At the same time, we excite the resonator indirectly by applying a square-envelope with frequency $f^{\prime}$ to XY-line of the test qubit by crosstalk interactions. Subsequently, we measure the populations of the ancilla qubit after a qubit-resonator swap interaction time $\pi/g^{\prime}$.
When the pulse frequency $f^{\prime}$ is close to $\omega_r(t)$, the resonator will be excited and reflected by the excitation of the ancilla qubit.
As shown in the Fig.~\ref{xidwc}b, the high-value populations show the trend of $f(t)$ and $\omega_r(t)$.
At the beginning of the experiment, the ancilla qubit stays at its idle frequency $f_0/2\pi=f(0)/2\pi=5.93$ GHz, where we define $\delta = \delta(0) = 0$. Therefore the effective resonator frequency $\delta(t)$ can be well controlled.

\subsection{Phase optimization of two longitudinal modulations}
To implement the effective Rabi Hamiltonian of Eq.~(1) in the main Text, the excitation energy of the test qubit is periodically modulated.
However, due to the imperfect waveform of the periodically modulated excitation energy $\omega_{q}t$, it's necessary to add an optimizable phase to each of the two modulated pulses, in our experiment, of the form
\begin{eqnarray}
    \omega_{q}(t) = \omega_{0} + \varepsilon_{1}\cos(\nu_{1}t+\phi_1)+ \varepsilon_{2}\cos(\nu_{2}t+\phi_2),
\end{eqnarray}
so as to validate the most appropriate dynamics of the Rabi Hamiltonian $H_{R}$.
For simplicity, we assume that $\phi_1$ and $\phi_2$ are independent of each other. During the dynamic process, $\varepsilon_2$ is slowly decreased. It is fairly assumed herein that the change in $\varepsilon_2$ has no impact on the value of $\phi_2$.

\begin{figure}[htb]
    \centering
    \includegraphics[width=14cm]{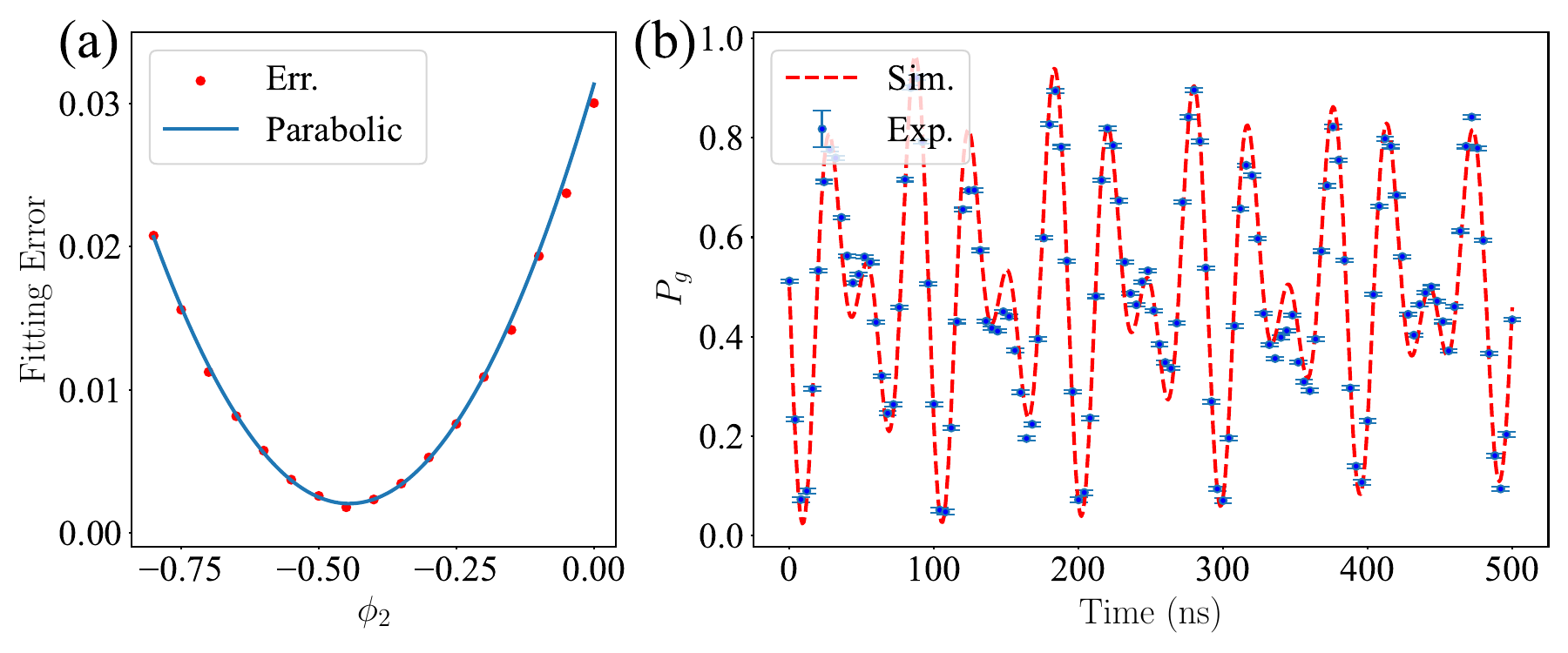}
    \caption{\textbf{Optimizing $\phi_2$ and population oscillation}. (a) Fitting errors versus the phase $\phi_2$ modification in the experiment. We compare populations with the correspondingly simulated oscillations by calculating the fitting error of the difference of the two populations (dots) and find the best $\phi_2$ (x-axis) by locating the minimum of the fitted parabolic (blue solid line). (b) Fitting for $\phi_1=0.0$ and $\phi_2=-0.45$. The populations $P_g$ of the ground state of the test qubit versus time. The blue dots and red dashed line represent the experimental data and the correspondingly numerical fitting results, respectively.}
    \label{phi2}
\end{figure}

The phase $\phi_1$ is optimized first. We apply the first frequency modulation pulse which has a fixed frequency $\nu_1/2\pi = 185$ MHz and amplitude $\varepsilon_1 = 2\pi \times 146$ MHz, and also a transverse microwave drive of an amplitude $A = 2\pi\times 15$ MHz, to the test qubit, initially in $(\vert g\rangle + \vert e \rangle)/\sqrt{2} $ state. The effective Hamiltonian is
\begin{eqnarray}\label{eq24}
    H_{\rm opt} = gJ_2(\mu)(\sigma^\dagger a +\sigma^-a^\dagger)+AJ_0(\mu)(\sigma^\dagger + \sigma^-).
\end{eqnarray}
We traverse $\phi_1$ between $-\pi$ and $\pi$, and measure the qubit population after 200 ns of the system evolution.
By comparing the population oscillation with the ideal numerical simulation results calculated through the evolution under the domination of the Hamiltonian of  Eq.~(\ref{eq24}), we obtain $\phi_0 = 0$.
Before optimizing $\phi_2$, we obtain $\nu_2=25.97$ MHz by fitting the population oscillation from the dynamics dominated by Eq.~(\ref{eq24}) (with the qubit's $\vert g \rangle$ state). We iterate over different phases $\phi_2$ to carry out the experiments with $\delta=0$ (by setting the ancilla qubit at its idle frequency) and $\varepsilon_2 = 10$ MHz. Finally, we obtain the one that best coincides with the correspondingly simulated Rabi oscillation curves. Fig.~\ref{phi2}a shows the values of the fitting error corresponding to different $\phi_2$. Here the fitting error is defined as $\sum (P^g_j-P_j^{ideal})^2/N$, where $N$ is the number of data points. Note that we choose the initial state $(\vert g\rangle - i \vert e \rangle)/\sqrt{2} $ instead of $\vert g\rangle $ as previously used because $\vert g\rangle $ is closer to the eigenstate of the system, and the effect on the population is relatively insignificant. We also slightly adjust the center frequencies and amplitudes of the two modulating pulses in order to improve the control. Based on the previous assumption, we choose the same phase modifications ($\phi_1$=0.0, $\phi_2$=-0.45) for different $\Omega$ throughout the experiment.
\section{Characterization of the resonator state}
\subsection{Meausrement of photon-number distribution}
\begin{figure}[htb]
    \centering
    \includegraphics[width=6cm]{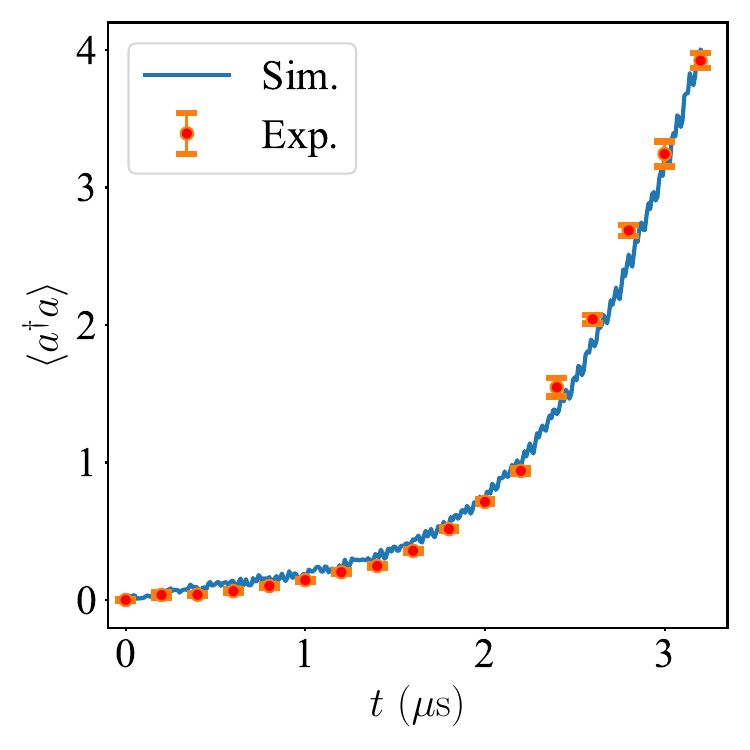}
    \caption{\textbf{Observed dynamical evolution of the average photon number $\bar{n}=\langle a^\dagger a \rangle$}. The red dots represent the average number of photons obtained from the population distribution in the experiment and the corresponding simulation result is shown by the blue curve.}
    \label{photon}
\end{figure}

All the Wigner function values in the main text are deduced from the photon-number distribution of the bus resonator. In the experiment, after a specific time $t$ of the SPTs dynamics evolution, the microwave drive is switched off. A displacement operation is applied to the resonator by a resonant pulse of 100 ns. Then, we bias the ancilla qubit from its idle frequency of 5.93 GHz to the frequency of 5.581 GHz where it is resonantly coupling with the resonator for a given time $\tau$.  In this case, the ancilla qubit undergoes photon-number-dependent Rabi oscillations. The populations $P^a_e(\tau)$, of the excited state of the ancilla qubit for a given interaction time $\tau$, are measured by biasing it back to its idle frequency, where the state readout is performed. The quantum Rabi oscillation signals can be fitted as
\begin{eqnarray}\label{Pnfit}
    P_{e}^{a}(\tau)=\frac{1}{2}\left[1-P_g^a(0)\stackrel{n_{\max }}{%
        \mathrel{\mathop{\sum }\limits_{n=0}}
    }P_{n}e^{-\kappa _{n}\tau}\cos \left(2\sqrt{n}\lambda' \tau\right)\right],
\end{eqnarray}
where $P_n$ denotes the $n$-photon state distribution probability, $n_{max}$ is the cutoff of the photon number in the fitting program, $P^a_g(0)$ indicates the probability that the ancilla qubit is initially in the ground state, and $\kappa_{n}=n^{l}/T_{1,p}$ ($l=0.7$) \cite{Pnfit1,Pnfit2,Pnfit3,Pnfit4,Pnfit5} is the empirical decay rate of the Rabi oscillations associated with the $n$-photon state. It is worth noting that due to the finite detuning $\Delta/2\pi = 0.392$ GHz between the ancilla qubit and the bus resonator, as the number of photons increases, the ancilla qubit inevitably interacts with the resonator during the SPTs dynamics, the ground state population $P^a_g(0)$ may not be $1$ at $\tau=0$. This induces a slight entanglement between the ancilla qubit and the resonator. Referring to \cite{Pnfit1}, we ignore the small excitation of $\vert e\rangle$ and  $P^a_g(0)$ is introduced to make the fitting more accurate. We then use the least square method to fit the experimental data according to Eq.~(\ref{Pnfit}) to obtain the actual photon number distribution, average photon number $\bar{n}=\langle a^\dagger a \rangle$, and further the Wigner function. Based on the measurements and the fitting operations described above, we can give the evolution of the average photon number in the resonator, as shown in the Fig.~\ref{photon}.
\subsection{Reconstruction of conditional Wigner functions}
As described in the main text, the resonator's Wigner function associated with the qubit state $\vert j\rangle~(j=e,g)$ is given by
\begin{eqnarray}
	{\cal W}_{j}(\beta)=\frac{2}{\pi P_j}\sum(-1)^n P_{j,n}(\beta),
\end{eqnarray}
where $P_{j,n}(\beta)=\langle j,n\vert D(-\beta)\rho D(\beta)\vert j,n\rangle $, and $P_{j}$ is the probability of detecting the qubit in $\vert j\rangle $-state. The photon number distribution $P_{j,n}(\beta)$ of the displaced field state associated with the state $\vert j\rangle $ of the test qubit can be extracted by resonantly coupling an ancilla qubit to the resonator, and correlating the observed Rabi signal to the measurement outcome $\vert j\rangle $ of the test qubit.
Here, to fit the photon-number-dependent Rabi oscillations better, we try to select the time point where the test qubit $\vert g \rangle $ and $\vert e\rangle $ states' populations are close to each other. With the increase of the photon number and thus also the expanded phase space, the fitting of the photon population becomes inaccurate. Some points with large fitting errors are neglected when we reconstruct the density matrix. A group of Wigner functions need to be continuously collected for 20 hours, and it is difficult to maintain qubit performance for a long time, so we again bring up qubits during the data collection process and the related data lacks error bars.

\begin{figure}[htb]
\centering
\includegraphics[width=18cm]{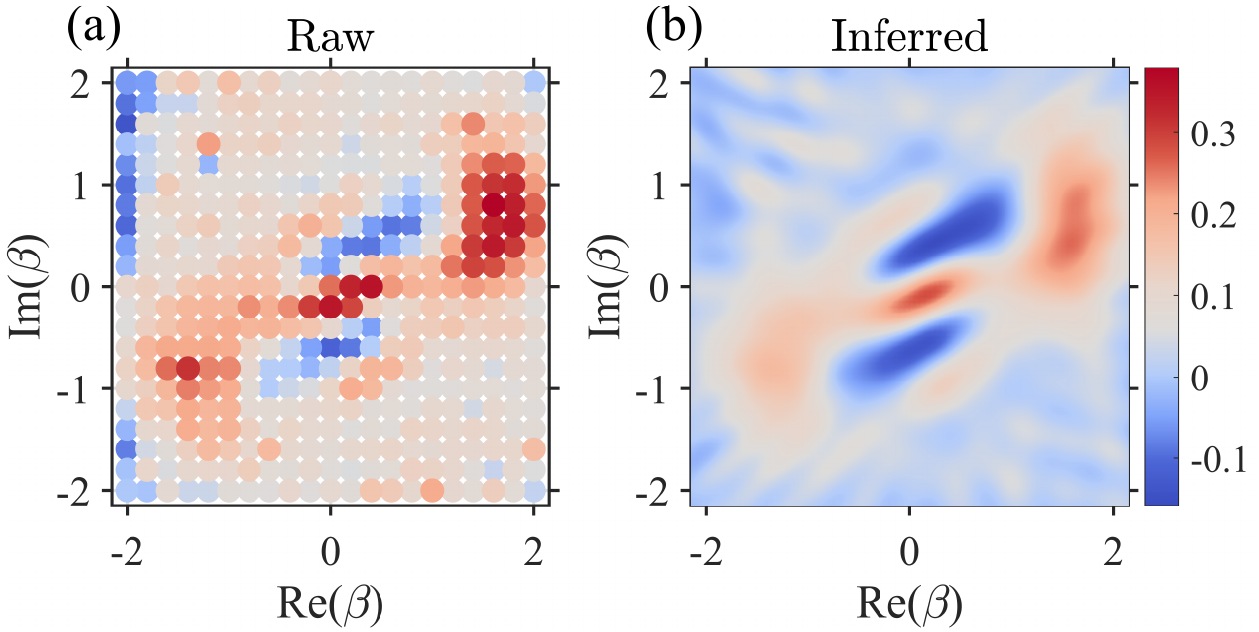}
\caption{\textbf{Comparison of raw and inferred Wigner functions.} (a), (e), (i)  Raw Wigner functions associated with the $\vert e \rangle $-state of the test qubit at 2.5, 2.8, and 3.0 $\mu$s.
(c), (g), (k)  Raw Wigner functions associated with the $\vert g \rangle $-state of the test qubit at 2.5, 2.8, and 3.0 $\mu$s.
(b), (d), (f), (h), (j), (l) Correspond inferred Wigner functions deduced from the raw data by CVX. }
\label{raw&inferred}
\end{figure}

After that, we use the CVX toolbox based on MATLAB \cite{cvx} to realize density matrix $\rho_{j,k}$ reconstruction from Wigner functions ${\cal{W}}_{j,k}$. The solved density matrix $\rho_{k,k}$ is Hermitian and positive semidefinite as well as satisfying ${\rm Tr}(\rho_{k,k})=1$. The raw Wigner data at $t=2.8$ $\mu$s and the corresponding Wigner functions calculated by the reconstructed density matrix are shown in Fig.~\ref{raw&inferred}, respectively. And we take the same treatment as that for Fig.~2 in the main Text.

\subsection{Correction of the phase-space rotations}\label{qfun}
For a more intuitive description of the phase space, we also use the Q-function to characterize the phase space distribution, defined as, $Q(\gamma )=\frac{1}{\pi }\left\langle \gamma \right\vert \rho _{r}\left\vert
\gamma \right\rangle $, where $\left\vert \gamma \right\rangle $ is the
coherent state with a complex amplitude $\gamma $ and $\rho_r$ is the reduced density operator of the resonator, which is obtained by reconstruction of the corresponding Wigner functions.
In the experiment, when the test qubit is in its $\vert e\rangle$ and $\vert g\rangle$ states, two factors cause the resonator states to rotate differently. One is due to a slight difference between the experimental frame and the effective Hamiltonian frame, and the other is generated by the resonator's Wigner tomography pulse.
Hence, after reconstructing the density matrix $\rho_k$ from the Wigner tomography $\mathcal{W}_k$, which is the reduced density operator of the resonator when the test qubits in $\vert k\rangle$ state, we add a rotation operator on the corresponding density matrix numerically. The rotation operator is defined as
\begin{eqnarray}\label{rotation}
U_k&=&\exp\left[-i\theta_k t/t_f a^\dagger a\right], \\
\rho_{k}^{\prime} &=& U_k \rho_k U_k^\dagger,\ \ \ \ \ \ \ \ \ \ (k=e,g)
\end{eqnarray}
and $\rho_r$ is obtained by adding two rotated $\rho_k^{\prime}$ according to the corresponding qubit state population:
\begin{eqnarray}\label{wigreson}
    \rho_r(t) = P_e^t\rho_e^{\prime}(t) + P_g^t\rho_g^{\prime}(t).
\end{eqnarray}
Here we choose $\left( \theta_e, \theta_g, t_f\right) = \left(10.6,10.2, 3.0 \ \mu {\rm s} \right)$ in the simulation.

The phase-space coordinates of the phase-space local maxima $\alpha_-$ and $\alpha_+$ in Fig.~1 in the main text was not well distinguished before $2$ $\mu$s, so we performed Gaussian fitting on the data, as shown by the dashed line in Fig.~1b.

\section{Numerical simulation}
\begin{figure}[htb]
    \centering
    \includegraphics[width=6cm]{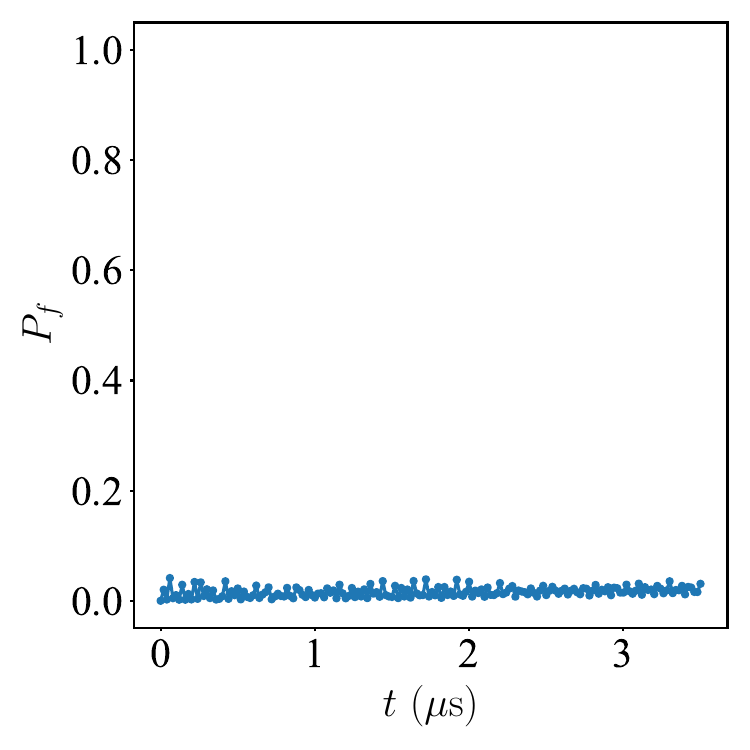}
    \caption{\textbf{Numerically simulated population of the third level $\vert f\rangle$ versus times.}
    }
    \label{pf}
\end{figure}

To further confirm the accuracy of the experimental results, we use the original Hamiltonian Eq.~(\ref{eqS1}) for numerical simulation for which the qubit's third level is included. We subtract the Stark term of the $a^\dagger a$ form to make the results more accurate, achieved experimentally by slightly adjusting the frequency of the auxiliary qubit. For simplicity, we directly shift the value of $\delta$ instead of introducing ancilla qubits, which is reasonable according to Fig.~\ref{xidwc}b. We plot the population of the third level $\vert f\rangle $ in Fig.~\ref{pf}, with the results showing that the average population of $\vert f \rangle$ is about 0.016 during the quenching dynamics, while the maximum population is 0.046. It can be seen that the second excited state $\vert f \rangle$ has little influence on the evolution of  the dynamics of the test qubit.

\begin{figure}[htb]
    \centering
    \includegraphics[width=17cm]{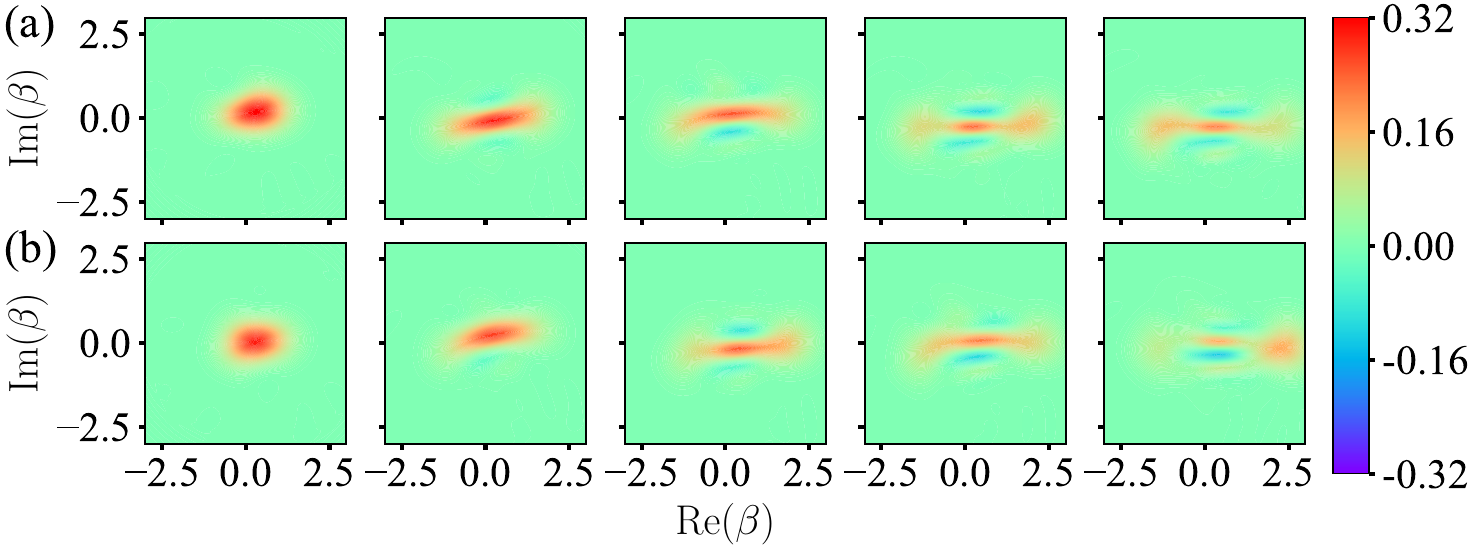}
    \caption{\textbf{Observation of the loss of the quantum coherence in numerical simulation}. (a) Conditional Wigner function ${\cal W}_{e}(\beta)$; (b) ${\cal W}_{g}(\beta)$ measured after quench times $t=$1.0, 2.0, 2.5, 2.8, and 3.0 $\mu$s.
    }
    \label{wigner}
\end{figure}

The simulated average photon number has been shown in Fig.~\ref{photon}, and the corresponding Wigner function of the diagonal elements are shown in Fig.~\ref{wigner}. These five columns show the Wigner functions at different times: 1.0, 2.0, 2.5, 2.8, and 3.0 $\mu$s.
